\documentclass[10pt,prd,superscriptaddress,amsfonts,amssymb,amsmath,showpacs,twocolumn]{revtex4-2}

\usepackage{amsmath,amssymb,amsfonts}
\usepackage{graphicx}
\usepackage{palatino}
\usepackage{mathpazo}
\usepackage{textcomp}
\usepackage{float}
\usepackage{booktabs}
\usepackage{dcolumn}
\usepackage{multirow}
\usepackage{xcolor}
\usepackage{hyperref}
\usepackage{orcidlink}

\hypersetup{
    colorlinks = true,
    linkcolor  = blue,
    citecolor  = blue,
    urlcolor   = blue
}

\begin{document}

\color{black}  

\title{Cosmological Evolution of Viscous Dark Energy in $f(Q,C)$ Gravity: Two-Fluid Approach}

\author{N. Myrzakulov\orcidlink{0000-0001-8691-9939}}
\email{nmyrzakulov@gmail.com}
\affiliation{L. N. Gumilyov Eurasian National University, Astana 010008, Kazakhstan}

\author{Anirudh Pradhan\orcidlink{0000-0002-1932-8431}}
\email{pradhan.anirudh@gmail.com}
\affiliation{Centre for Cosmology, Astrophysics and Space Science (CCASS), GLA University, Mathura-281406, U.P., India}

\author{S. H. Shekh\orcidlink{<correct-ORCID-here>}} 
\email{da\_salim@rediff.com}
\affiliation{L. N. Gumilyov Eurasian National University, Astana 010008, Kazakhstan}
\affiliation{Department of Mathematics, S.P.M. Science and Gilani Arts, Commerce College, Ghatanji, Yavatmal, Maharashtra-445301, India}

\author{Anil Kumar Yadav\orcidlink{0000-0002-5174-5542}}
\email[Corresponding author: ]{abanilyadav@yahoo.co.in}
\affiliation{Department of Physics, United College of Engineering and Research, Greater Noida - 201310, India}

\begin{abstract}
\textbf{Abstract:}  
In this paper, we explore the cosmological evolution of a viscous dark energy model within the framework of $f(Q, C)$ gravity, employing a two-fluid approach. The model incorporates non-metricity and boundary contributions to the total action, represented by the scalar quantities $Q$ and $C$. The viscosity in the dark energy fluid is introduced to investigate the impact of bulk viscosity on cosmic expansion and late-time acceleration. Field equations are derived in a modified FLRW background, and the dynamics of key cosmological quantities such as energy density, pressure, and the effective equation of state (EoS) parameter are analyzed. Observational constraints on $H(z)$ are imposed using DESI BAO Measurements, Pantheon+ (without SHOES), and cosmic chronometer datasets. Results indicate that the model effectively captures the universe’s expansion history, including the deceleration–acceleration transition, consistent with observations. This framework provides an alternative explanation for late-time cosmic acceleration without invoking a cosmological constant. \\

\textbf{Keywords:} FRW universe; effective equation of state; $f(Q,C)$ gravity; cosmology.
\end{abstract}

\maketitle

\section{Introduction}\label{I}

In the past decade, cosmological studies have indicated that Dark Energy (DE) is the dominant component in the current universe, leading to its accelerated expansion \textcolor{blue}{(Riess (1998), Perlmutter (1999), Koivisto (2006), Daniel (2008), Bennett (2003), Spergel (2003), Hinshaw (2013), Caldwell (2004), Huang (2006), Eisenstein (2005)), Percival (2010))}. Observational evidence from various cosmological datasets shows that the universe is spatially flat, with approximately 70\% of its content being DE, about 30\% consisting of Cold Dark Matter (CDM) and baryons, and only a small fraction represented by radiation. Despite knowing that DE significantly influences the universe's long-term evolution, its fundamental properties and origin remain a mystery. Numerous models have been put forward to explain the characteristics of DE, including some that propose dynamic behavior. Among these models are the evolving canonical scalar field theory known as quintessence, characterized by an Equation of State (EoS) parameter in the range $-1 < \omega < -\frac{1}{3}$; phantom energy, with an EoS parameter $\omega < -1$ which violates the Weak Energy Condition (WEC); and quintom energy, in which the EoS transitions through $\omega = -1$  \textcolor{blue}{(Ratra (1998), Sami (2004), Armendariz (2000), Khoury (2004), Padmanabhan (2002), Bento (2002), Zarrouki (2010), Elizalde (2004), Nojiri (2003))}. The current value of the EoS parameter for DE has been derived from a combination of WMAP9 (Nine-Year Wilkinson Microwave Anisotropy Probe) data, along with Hubble parameter measurements ($H_0$), Type Ia Supernovae (SNIa), the Cosmic Microwave Background (CMB), and Baryon Acoustic Oscillations (BAO). These datasets suggest that the present-day value of $\omega_0$ is approximately $-1.084 \pm 0.063$  \textcolor{blue}{(Hinshaw (2013))}. The Planck collaboration further refined this value in 2015 to $\omega_0 = -1.006 \pm 0.0451$  \textcolor{blue}{(Ade (2015)) } and the 2018 results gave a more precise estimate of $\omega_0 = -1.028 \pm 0.032$  \textcolor{blue}{(Ade (2015))}.

In recent years, several innovative approaches have been proposed as alternatives to the conventional Einstein-Hilbert action, which is the foundation of General Relativity, in order to address the challenge of explaining the universe's accelerating expansion. These alternative approaches are collectively known as Modified Theories of Gravity (MTG). Under MTG, many different actions have been formulated to account for this cosmic acceleration. Some of the most widely studied modifications include:
	
\textbf{\textit{$f(R)$ Gravity}}: This theory extends General Relativity by generalizing the Ricci scalar $R$ in the Einstein-Hilbert action to a more general function $f(R)$. This extension is designed to explain cosmic acceleration without the need for dark energy  \textcolor{blue}{(Capozziello (2011),Sotiriou (2010),Nojiri (2010),Nojiri (2017),Brevik (2017)}.
	
\textbf{\textit{$f(T)$ Gravity}}: $f(T)$ gravity modifies the Teleparallel Equivalent of General Relativity (TEGR), replacing the torsion scalar $T$ with a function $f(T)$. In this framework, torsion rather than curvature is responsible for describing gravitational effects  \textcolor{blue}{(Ferraro (2007), Krssak (2019))}.
	
\textbf{\textit{$f(G)$ Gravity}}: $f(G)$ gravity incorporates a general function of the Gauss-Bonnet invariant $G$, which includes contributions from both the Ricci scalar and the Riemann curvature tensor. This theory attempts to provide insights into gravitational interactions and cosmic acceleration  \textcolor{blue}{(Nojiri (2005), Bamba (2012))}.
	
\textbf{\textit{$f(Q)$ Gravity}}: This model uses the non-metricity scalar $Q$ to explain gravitational phenomena. It is a part of metric-affine geometry, where gravity is described through variations in the length of vectors rather than their direction, making it distinct from other geometric approaches  \textcolor{blue}{(Jimenez (2018),D'Ambrosio (2019), Heisenberg (2023))}.
	
\textbf{\textit{$f(Q,T)$ Gravity}}: $f(Q,T)$ gravity extends $f(Q)$ gravity by incorporating the trace of the energy-momentum tensor $T$ in the action, enabling an interaction between matter and geometry. This coupling introduces additional flexibility in modeling cosmic acceleration  \textcolor{blue}{(Harko (2018), Anagnostopoulos (2020))}.
	
Furthermore, the theory known as \textbf{\textit{$f(Q,C)$ Gravity}} has been introduced to provide new insights into the nature of dark energy and the accelerating expansion of the universe. By allowing for nonlinear dependencies on $Q$, this theory explores potential explanations for late-time cosmic acceleration without requiring exotic fields or a cosmological constant. The inclusion of the boundary term $C$ introduces novel gravitational effects, which can be tested through observational data, such as those from the cosmic microwave background (CMB), large-scale structure (LSS), and type Ia supernovae (SNIa).
	
In $f(Q,C)$ gravity, the non-metricity scalar $Q$ quantifies the deviation of the metric from being preserved during parallel transport. Unlike General Relativity, where the connection is torsion-free and symmetric, $f(Q,C)$ also incorporates the boundary term $C$, which arises due to the interplay between torsion-free and curvature-free connections. This boundary term ensures the theory remains dynamically equivalent to General Relativity under special conditions, allowing for smooth transitions between different geometric descriptions of gravity. The term $C$ provides additional degrees of freedom, influencing the behavior of gravitational fields, particularly on cosmological scales. One of the key motivations for developing $f(Q,C)$ gravity is the unification of different geometric frameworks for gravity, including curvature-based, torsion-based, and non-metricity-based theories. By incorporating both the non-metricity scalar $Q$ and the boundary term $C$, the theory offers a unified framework that can interpolate between Teleparallel Gravity, General Relativity, and other modified gravity theories. \\

The gravitational action of the $f(Q, C)$ gravity theory is given by
	
	\begin{equation}\label{e1}
		S=\int \bigg(\frac{1}{2k}f(Q,C)+\mathcal{L}_{m}\bigg)\sqrt{-g}d^{4}x,
	\end{equation}
	The field equation can be formally derived by performing a metric variation of the action presented in equation (\ref{e1}), which subsequently yields:
	\begin{small}
		\begin{widetext}
			\begin{equation}\label{e2}
\kappa T_{\mu\nu}=-\frac{f}{2}g_{\mu\nu}+\frac{2}{\sqrt{-g}}\partial_{\alpha}\bigg(\sqrt{-g}f_{Q}P^{\alpha}_{\mu\nu}\bigg)+\bigg(P_{\mu\eta\beta}Q_{\nu}^{\eta\beta}-2P_{\eta\beta\nu}Q^{\eta\beta}{\mu}\bigg)f{Q}+\bigg(\frac{C}{2}g_{\mu\nu}-\overset{\circ}\nabla_{\mu}\overset{\circ}{\nabla_{\nu}}+g_{\mu\nu}\overset{\circ}{\nabla^{\eta}}\overset{\circ}\nabla_{\eta}-2P^{\alpha}{\mu\nu}\partial{\alpha}\bigg)f_{C},
			\end{equation}
		\end{widetext}
	\end{small}

The development of $f(Q,C)$ gravity stems from the desire to broaden gravitational theories to offer new interpretations for cosmic acceleration while uniting various geometric frameworks. This approach is consistent with recent studies, such as the work by Jimenez et al.  \textcolor{blue}{(2021)}, which explores the geometric trinity of gravity. Their research outlines how curvature, torsion, and non-metricity provide alternative insights into gravitational theory. Building upon these concepts, $f(Q,C)$ gravity incorporates both the non-metricity scalar and boundary terms to extend the standard models of gravity. Frusciante (2021) further examines the implications of this framework in a cosmological setting, emphasizing the significant role of the boundary term $C$ in influencing gravitational interactions. His work also identifies specific observational signals that could differentiate $f(Q,C)$ gravity from other existing models. In another study, Zhao and Cai  \textcolor{blue}{(2021)} investigate the dynamic properties of $f(Q,C)$ gravity with a focus on its impact on cosmological evolution. They highlight the boundary term's contribution to addressing the universe's accelerating expansion. Anagnostopoulos et al.  \textcolor{blue}{(2022)} delve into the theory's stability and its cosmological implications, particularly regarding how the boundary term affects the universe's evolution and its potential to provide an explanation for dark energy.

De et al. \textcolor{blue}{(2024) } recently developed $f(Q,C)$ gravity and cosmology by including the boundary term $C$ in the Lagrangian along with $Q$. After extracting the general field equations, they applied them to the flat Friedmann-Robertson-Walker (FRW) metric inside a cosmological framework. By summing $f(Q,C) = \alpha Q^{n} + \beta C$, the rip cosmology theories of the Universe have been given for the $f(Q,C)$ gravity theory and the nature of the physical parameters for the Little Rip, Big Rip, and Pseudo Rip models is analyzed by Samaddar et al.  \textcolor{blue}{(2024)}. In order to better understand the function of the boundary term in $f(Q, B)$ symmetric teleparallel gravity, Capozziello et al. \textcolor{blue}{(2023)} also examined the Gibbons-Hawking-York boundary term of General Relativity and contrasted it with $B$ in $f(Q,C)$ gravity. In $f(Q,C)$ gravity theory, Maurya  \textcolor{blue}{(2024, 2024a, 2024b, 2024c)} studied an isotropic and homogeneous flat dark energy model that is linear in non-metricity Q and quadratic in boundary term C as $f(Q,C)=Q+ \alpha C^2$, where $\alpha$ is a model parameter. Using MCMC analysis, he (Maurya 2024) compared the derived results with two observational datasets, $H(z)$ and Pantheon SNe Ia datasets, and they found the best fit current values of parameters.

In a flat FLRW spacetime universe, Pradhan et al.  \textcolor{blue}{(2024)} have proposed a modified non-metricity gravity theory with boundary terms, this study examines dark energy scenarios of cosmological models with observational constraints by taking into account the arbitrary function $f(Q,C)=Q+\alpha C^2$.

Motivated by these foundational works, we aim to expand this area of research by performing an in-depth analysis within the $f(Q,C)$ gravity framework.	This makes $f(Q,C)$ gravity a promising candidate for investigating the geometric structure of spacetime and its influence on cosmic evolution.

	
\section{Metric and field equations}\label{II}
This paper considers a homogeneous and isotropic universe, described by the FLRW spacetime with the following form:
\begin{equation}\label{e3}
	ds^{2}=dt^{2}-a^{2}(t) dr^{2}-a^{2}(t)r^{2} d\theta ^{2}-a^{2}(t)r^2\sin^{2}\theta d\phi ^{2},  
\end{equation}
where $a(t)$ represents the universe scalar factor which is dependent on the cosmic time $t$,  and $\left( t,r,\theta,\phi\right)$ denotes the comoving coordinates. As thus, the non-metricity tensor trace is
\begin{equation}\label{e4}
	Q=6H^{2}. 
\end{equation}
The stress-energy tensor is provided by the following when we consider the matter to be a perfect fluid:
\begin{equation} \label{e5}
	T_{\mu \nu }=\left( \rho +\bar{p}\right) u_{\mu }u_{\nu }-\bar{p}g_{\mu \nu },  
\end{equation}
where $\bar{p}=p-3\xi H$, the four-velocity is $u_{\mu }$ which follows $u_{\mu }u^{\mu }=1$, the  energy density is $\rho $ and isotropic pressure is $p$. 
By employing equations (\ref{e3}) and (\ref{e5}), we formally derive the field equations, which take the form:
\begin{equation} \label{e6}
	\rho=6H^2f_Q-(9H^2+3\dot{H})f_C+3H\dot{f}_C+f/2 
\end{equation}

\begin{equation} \label{e7}
	\bar{p}=-(6H^2+2\dot{H})f_Q-2H\dot{f}_Q+(9H^2+3\dot{H})f_C-\ddot{f}_C-f/2 
\end{equation}
where the overhead dot represent the differentiation with cosmic time $t$. $H$ be the Hubble parameter, $f_Q=\frac{\partial f}{\partial Q}$ and $f_C=\frac{\partial f}{\partial C}$. The field equations (\ref{e6}) and (\ref{e7}) exhibit nonlinear behavior, rendering their solutions challenging to obtain. To address this, we explore a nonlinear $f(Q, C)$ gravity model of the form:
\begin{equation} \label{e8}
	f(Q,C)=a_1 Q^\gamma +a_2 C
\end{equation}

where $a_1>0$ and $a_2>0$ are the model parameters. The above model offers a promising extension of General Relativity, integrating non-metricity and boundary contributions. This framework flexibly addresses late-time cosmic acceleration without relying exclusively on dark energy or a cosmological constant. Incorporating the non-metricity scalar $Q$ and boundary term $C$ enhances the model's versatility, enabling geometric interpretations of cosmic acceleration and compatibility with diverse cosmological observations. The power-law $Q$ dependence and inclusion of $C$ facilitate adaptable data fitting and testable deviations from the $\Lambda$CDM paradigm, ensuring consistency with existing and future experiments  \textcolor{blue}{(Lazkoz (2019), Jimenez (2018),  Dialektopoulos (2021), Zhao (2022))}. By implementing the model (Eq. \ref{e8}), the field equations (Eqs. \ref{e6}, \;\ref{e7}) can be reformulated as:
\begin{widetext}
\begin{equation} \label{e9}
	\rho = \frac{1}{2} \left(6^{\alpha } a_1 \left(-H^2\right)^{\alpha }+6 a_2 \left(3 H^2+\dot{H}\right)\right)+6^{\alpha } \alpha   a_1  H^2 \left(-H^2\right)^{\alpha -1}- a_2 \left(9 H^2+3 \dot{H}\right)-\rho_0 (z+1)^{3 (-\sigma +\omega_ m+1)}
\end{equation}

and
\begin{equation} \label{e10}
	\bar{p} = 6^{\gamma -1}a_1 (2 \gamma -1) (-H)^{\gamma } H^{\gamma -2} \left(3 H^2+2 \gamma  \dot{H}\right)
\end{equation}
\end{widetext}
The corresponding equation of state parameters are molded as $\omega=\frac{\bar{p}} {\rho}$. \\
However, within the realm of cosmological theory, a novel paradigm has emerged, focusing on the dynamic interplay between the matters (dark matter and dark energy). This conceptual framework initially arose to reconcile the cosmological constant's unexpectedly low value but subsequently revealed its potential in resolving the enigmatic synchronization of DM and DE densities at the present epoch, known as the cosmic coincidence problem  \textcolor{blue}{(Bolotin (2015), Wang (2016))}. The Current observational evidence suggests a preference for interaction between dark matter and dark energy at late times   \textcolor{blue}{(Salvatelli (2014), Nunes (2016))}. Moreover, estimates of the coupling parameter in the dark sector have been obtained through various observational datasets  \textcolor{blue}{(Xia (2016), Caprini (2016))}. Recent studies have extensively explored various interacting dark energy models. Notably, current observational data suggest a nonzero interaction between dark sectors, hinting at novel possibilities. This promising avenue warrants further investigation, potentially unveiling new insights into the dark universe.\\
The interacting dynamics allow dark matter and dark energy to exchange energy and momentum as the universe evolves, the conservation equations for dark matter and dark energy are $\Delta^\nu T^{m}_{\mu\nu}=\mathcal{Q}$ and  $\Delta^\nu T^{\Lambda}_{\mu\nu}=-\mathcal{Q}$. Hence, The resulting conservation equations for dark matter and dark energy are:

\begin{equation} \label{e11}
\dot{\rho}_m +3H\rho_m=\mathcal{Q} {\;\;\;}\text{and}{\;\;\;} \dot{\rho}_{\Lambda} +3H(\rho_{\Lambda}+\bar{p})=-\mathcal{Q}
\end{equation}
Equation (\ref{e11}) introduce a novel cosmic dynamics, incorporating the interaction function $\mathcal{Q}$. Despite numerous proposed interactions in the literature, the precise form of $\mathcal{Q}$ remains elusive. The coupled continuity equations reveal that $\mathcal{Q}$ can be an arbitrary function, 
enabling the construction of countless interacting models. We choose the following interaction as a starting point:
\begin{equation} \label{e12}
\mathcal{Q}=3H\sigma \rho_m
\end{equation}
where $\sigma$ be the coupling constant. Using the above equation (\ref{e12}) in equation (\ref{e11}) and after integration we obtain $\rho_m=\rho_0 (1+z)^{3(1+\omega_m-\sigma)}$. To simplify our analysis and justify model parameters, we need to determine the Hubble parameter ($H$). To this end, in this study focuses on a specific parametrization of the deceleration parameter ($q$), given by
\begin{equation} \label{8}
q=-1+\frac{\alpha}{1+\beta (1+z)^{-3}}
\end{equation}
The parametrization of the deceleration parameter adopted in this work is motivated by its ability to interpolate smoothly between a decelerating matter-dominated phase at high redshifts and an accelerating phase at late times. Explicitly, the form of $q(z)$ ensures $q>0$ for $z \gg 1$ and $q<0$ at $z \rightarrow 0$, thereby reproducing the expected cosmic history in a minimal two-parameter description.\\	
Although the $\Lambda$CDM model corresponds to a specific limiting case of this parametrization, the present approach allows for deviations that can be directly constrained by observations. In particular, the closed-form expression of $H(z)$ obtained from this parametrization is advantageous for confronting the model with data such as DESI, Pantheon+, and Cosmic Chronometers. Thus, the parametrization is not merely a rephrasing of $\Lambda$CDM but a controlled generalization that accommodates dynamical features consistent with viscous $f(Q,C)$ gravity.

The latest observational data from SNe Ia and CMB point to accelerating cosmic expansion, characterized by a negative deceleration parameter ($q$). At $z = 0$, our expression simplifies to $q_0 = -1 + \frac{\alpha}{1+\beta}$. This implies that the current cosmic dynamics are governed by the relationship between $\alpha$ and $\beta$, with three possible scenarios:
\begin{itemize}
	\item  $\alpha < 1+\beta:$ acceleration,
	\item  $\alpha > 1+\beta:$ deceleration,
	\item  $\alpha = 1+\beta:$ steady expansion.
\end{itemize}
By utilizing the equation linking $H$ and $q$ as $H=H_0exp\left( \int_{0}^{z}\frac{1+q}{1+z}dz\right)$, we derive an expression for the Hubble parameter $H(z)$ as follows.
\begin{equation} \label{9}
	H=H_0\left(\frac{(1+z)^3+\beta}{1+\beta}\right)^{\frac{\alpha}{3}}
\end{equation}
Here we explore the combined effects of various cosmological components rather than isolating a single factor. Specifically, the model integrates the nonlinear non-metricity scalar $Q$, the boundary term $C$, bulk viscosity, interaction between dark matter and dark energy, and a parametrized deceleration parameter $q$. Each component serves a distinct purpose: the $f(Q,C)$ framework introduces a geometric foundation for modified gravity, with $Q$ and $C$ capturing deviations from standard metricity and curvature paradigms, respectively. Viscosity represents dissipative processes in the cosmic fluid, which can influence the pressure and enhance late-time acceleration, while the interaction term captures potential energy-momentum exchanges between dark matter and dark energy. The deceleration parameter $q$, parametrized phenomenologically, is employed to reconstruct the Hubble parameter for empirical validation and parameter constraints. Together, these elements allow a comprehensive investigation into cosmic acceleration, understanding into how geometric, dynamic, and phenomenological factors collectively contribute to the universe's evolution. By emphasizing their interplay rather than individual dominance, we present a holistic perspective on accelerating cosmic expansion, consistent with theoretical and observational frameworks.

\section{Datasets and Statistical Analysis}
In order to constrain the free parameters of the viscous $f(Q,C)$ cosmological model, we employ three different classes of cosmological observations: (i) Cosmic Chronometers (CC), (ii) Baryon Acoustic Oscillations (BAO) from DESI, and (iii) Type Ia Supernovae (SNe). For each dataset, we construct the corresponding $\chi^2$ function, and the total likelihood is obtained by combining them in a joint analysis.

\noindent \textbf{Cosmic Chronometers (CC):} The cosmic chronometer technique provides direct, model-independent measurements of the Hubble parameter $H(z)$ using the differential age evolution of passively evolving galaxies \textcolor{blue}{Jimenez (2002), Moresco (2012)}. We use the compilation of CC data points available in \textcolor{blue}{Moresco (2016), Scolnic (2018), Moresco (2022)}.\\

\noindent \textbf{DESI BAO Measurements:} For the DESI Y1 BAO dataset, the primary observables are the comoving angular diameter distance $D_M(z)$, the Hubble distance $D_H(z)$, and their ratios with respect to the sound horizon $r_d$. They are defined as
\begin{widetext}
\begin{equation}
D_M(z) = (1+z) \int_0^z \frac{dz'}{H(z')}, 
\qquad 
D_H(z) = \frac{c}{H(z)}, 
\qquad 
r_d = \int_{z_d}^{\infty} \frac{c_s(z)}{H(z)} \, dz,
\end{equation}
\end{widetext}
where $c_s(z)$ is the baryon-photon sound speed.  \\
The DESI likelihood is constructed from the data vector $\mathbf{X} = \{ D_M(z)/r_d, D_H(z)/r_d \}$, and the $\chi^2$ is written as
\begin{equation}
\chi^2_{\rm DESI} = \Delta \mathbf{X}^{T} \, C^{-1} \, \Delta \mathbf{X},
\end{equation}
where $\Delta \mathbf{X} = \mathbf{X}_{\rm th} - \mathbf{X}_{\rm obs}$ and $C$ is the covariance matrix supplied by the DESI collaboration \textcolor{blue}{(Adame 2024)}. The inclusion of the full covariance is essential, since $D_M$ and $D_H$ at the same redshift are correlated. \\ 
In our analysis, $r_d$ is treated as a free parameter and is constrained simultaneously with the other cosmological parameters in the MCMC. The effective redshifts $z_{\rm eff}$ and the corresponding BAO measurements are listed in Table~\ref{tab:desi}, while the covariance matrix is taken directly from the DESI data release.
\begin{table}[H]
\centering
\caption{DESI Y1 BAO measurements \textcolor{blue}{(Adame 2024)}. The quoted errors are $1\sigma$ uncertainties. The full covariance matrix between $D_M/r_d$ and $D_H/r_d$ at each redshift is included in the likelihood.}
\label{tab:desi}
\begin{tabular}{c c c}
\hline
$z_{\rm eff}$ & $D_M(z)/r_d$ & $D_H(z)/r_d$ \\
\hline
0.51 & $13.92 \pm 0.24$ & $20.45 \pm 0.51$ \\
0.71 & $16.93 \pm 0.26$ & $19.65 \pm 0.45$ \\
0.93 & $20.45 \pm 0.31$ & $17.21 \pm 0.44$ \\
1.32 & $27.58 \pm 0.41$ & $13.26 \pm 0.33$ \\
1.49 & $30.12 \pm 0.52$ & $12.34 \pm 0.39$ \\
\hline
\end{tabular}
\end{table}

\noindent \textbf{Type Ia Supernovae (SN)}
We further include the Pantheon+ compilation of Type Ia supernovae \textcolor{blue}{Scolnic (2018), Brout (2022)}, which provides measurements of the distance modulus
\begin{equation}
\mu_{\rm th}(z) = 5 \log_{10} \left[ \frac{D_L(z)}{\text{Mpc}} \right] + 25,
\end{equation}
where the luminosity distance is
\begin{equation}
D_L(z) = (1+z) \int_0^z \frac{c \, dz'}{H(z')}.
\end{equation}


	

We performed a joint cosmological analysis combining Observational Hubble Data (OHD), Baryon Acoustic Oscillations (BAO), and Type Ia Supernovae (Pantheon +) to constrain a phenomenological expansion model of the Universe $H(z) = H_0 \left[ \frac{(1+z)^3 + \beta}{1+\beta} \right]^{\alpha/3}$ where \(H_0\) is the present-day Hubble constant, \(\alpha\) and \(\beta\) are expansion parameters, \(r_d\) is the sound horizon scale, and \(M\) is the absolute magnitude of SNe Ia. Gaussian priors on \(H_0\) and \(r_d\) from Planck were included to reduce degeneracies.

The total $\chi^2$ function is read as
\begin{equation}
\chi^2_{\rm total} = \chi^2_{\rm OHD} + \chi^2_{\rm BAO} + \chi^2_{\rm SN} + \chi^2_{\rm priors}.
\label{eq:chi2total}
\end{equation}

The contributions from each dataset are defined as follows.
\subsection{Observational Hubble Data (OHD)}
\begin{equation}
\chi^2_{\rm OHD} = \sum_{i=1}^{N_{\rm OHD}} 
\frac{\left[ H_{\rm obs}(z_i) - H_{\rm model}(z_i; H_0, \alpha, \beta) \right]^2}{\sigma_H(z_i)^2}.
\label{eq:chi2OHD}
\end{equation}
\subsection{Baryon Acoustic Oscillations (BAO)}
\begin{equation}
\chi^2_{\rm BAO} = \sum_{i=1}^{N_{\rm BAO}} 
\frac{\left[ D_V^{\rm model}(z_i; H_0, \alpha, \beta, r_d)/r_d - D_V^{\rm obs}(z_i)/r_d \right]^2}{\left(\sigma_{D_V}(z_i)/r_d\right)^2}.
\label{eq:chi2BAO}
\end{equation}
\subsection{Type Ia Supernovae (Pantheon-like)}
\begin{equation}
\chi^2_{\rm SN} = \sum_{j=1}^{N_{\rm SN}} 
\frac{\left[ \mu_{\rm obs}(z_j) - \mu_{\rm th}(z_j; H_0, \alpha, \beta, M) \right]^2}{\sigma_\mu(z_j)^2}.
\label{eq:chi2SN}
\end{equation}
\subsection{Gaussian Priors on $H_0$ and $r_d$}
\begin{equation}
\chi^2_{\rm priors} = \frac{(H_0 - H_0^{\rm prior})^2}{\sigma_{H_0}^2} + \frac{(r_d - r_d^{\rm prior})^2}{\sigma_{r_d}^2}.
\label{eq:chi2priors}
\end{equation}
We minimized \(\chi^2_{\rm total}\) using the L-BFGS-B algorithm with physically motivated bounds, and generated Monte Carlo samples assuming Gaussian errors to estimate parameter uncertainties. The resulting best-fit parameters are obtained as
\begin{align}
H_0 &= 67.1 \pm 1.6 \ \mathrm{km/s/Mpc}, \\
\alpha &= 1.51^{+0.18}_{-0.21}, \\
\beta &= 1.367 \pm 0.041, \\
r_d &= 146.9 \pm 3.4 \ \mathrm{Mpc}, \\
M &= -19.465 \pm 0.051.
\end{align}

\noindent Fig. 1 shows 1D marginalized distributions and 2D confidence contours, highlighting correlations among parameters. Inclusion of OHD significantly improves constraints on \(H_0\) and \(\alpha\), while the combination with BAO and SNe ensures tighter bounds on \(\beta\), \(r_d\), and \(M\).


\begin{widetext}
\begin{figure}[H]
	\centering
	\includegraphics[scale=1.2]{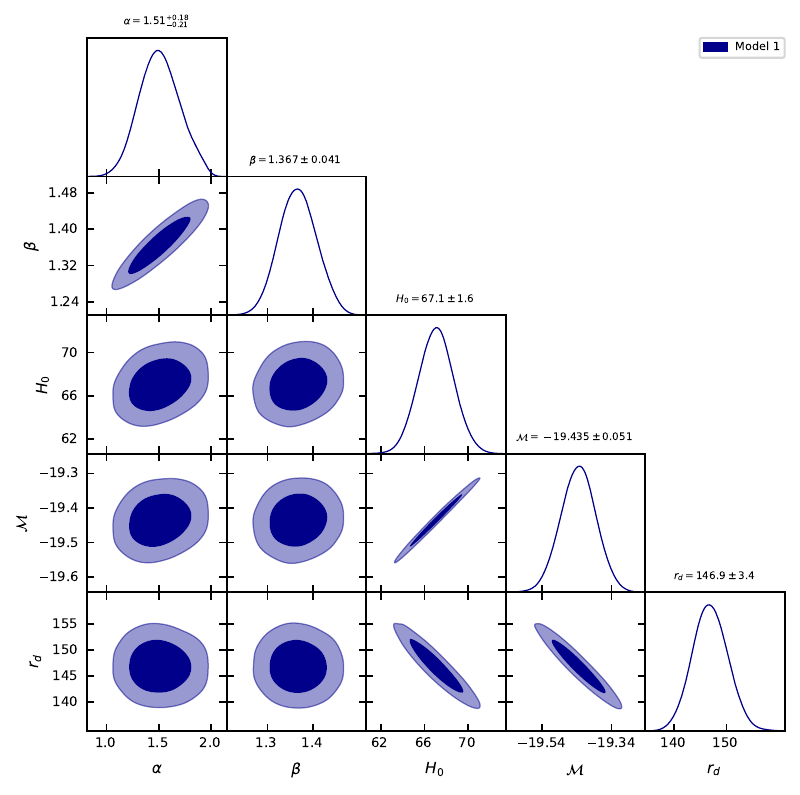}
	\caption{The likelihood contours for the model parameters, shown as $1\sigma$ and $2\sigma$ errors for the free parameters, are determined using the combined DESI-Y1, SDSS-IV, Pantheon$+$ (without SHOES) and Cosmic chronometer.}\label{MCMC}
\end{figure}
\end{widetext}
In the context of Fig. \ref{MCMC}, the parameters $M$ and $r_d$ hold specific roles in the analysis of cosmological datasets. The parameter $M$ refers to the absolute magnitude of Type Ia supernovae, which serves as a critical standard candle for determining luminosity distances and constraining the Hubble parameter $H(z)$. Accurate calibration of $M$ is essential for connecting observed supernova brightness to cosmological models. The parameter $r_d$ represents the comoving sound horizon at the baryon drag epoch, a standard ruler derived from Baryon Acoustic Oscillations (BAO) data, which provides scale information for the large-scale structure of the universe and is directly related to the measurement of expansion history.
\section{Cosmic physical parameters}

Our study focuses on exploring the influence of modified gravity theories, particularly incorporating viscosity, on the universe's cosmological parameters. By grounding the investigation in physical principles, the study aims to provide insights into how these modifications affect the following key parameters.\\
By using best-fit values of the model parameters, the study ensures that these modified models are consistent with observations, offering a scientifically justified perspective on cosmic evolution. The results could help bridge gaps in our understanding of dark energy, dark matter, and the accelerating expansion of the universe, enhancing our comprehension of fundamental cosmological processes.\\
\textbf{Energy Density:} Examines how modified gravity theories and viscosity contribute to the distribution and evolution of energy density in the universe. The understanding of energy density is crucial for the dynamics of cosmic expansion.
In our analysis, the energy density is a crucial parameter that reflects the total matter content of the universe at various redshifts, which correspond to different points in cosmic history.
\begin{figure}[H]
\centering
\includegraphics[scale=0.7]{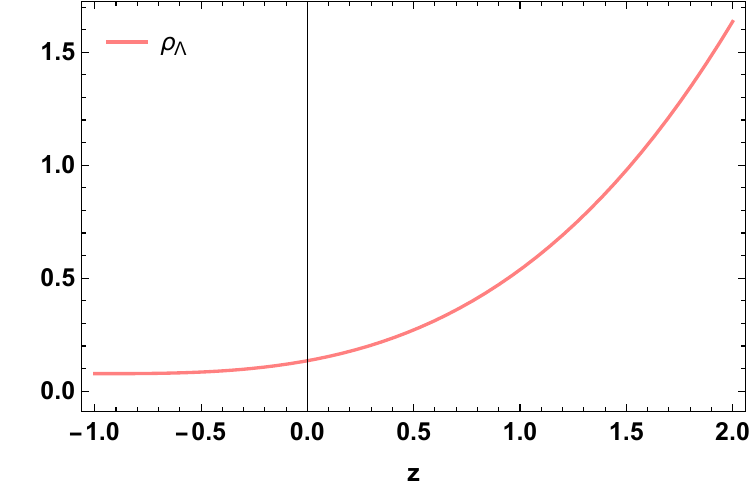}
\caption{The behavior of energy density of the fluid with the values of free parameters constraint by the combined data sets of DESI-Y1, SDSS-IV, Pantheon$+$ (without SHOES) and CC.}\label{den}
\end{figure}
From Fig. (\ref{den}), the energy density is shown to remain positive for the constrained values of the model parameters. This is important as it signifies that the model adheres to physically realistic conditions, where the universe maintains a significant amount of energy density, particularly in earlier epochs. As the universe expands, the energy density decreases. This aligns with the expectation that as space itself stretches, the matter (and hence energy density) dilutes over time. The energy density is observed to approach zero as \( z \) approaches \( -1 \) (the far future of the universe). This suggests that, according to your viscous \( f(Q, C) \) model, the universe might continue expanding at an accelerated rate, with the energy density gradually vanishing in the distant future. The behavior described aligns well with predictions from the standard \( \Lambda \)-Cold Dark Matter (\( \Lambda \text{CDM} \)) model. In both models, the energy density decreases over time, reinforcing the validity of your modified viscous gravity model. Hence, this finding strengthens the case for the viscous \( f(Q, C) \) cosmological model by showing that it can reproduce known results of cosmic expansion, such as the gradual dilution of energy density, while potentially offering additional insights into the effects of viscosity on cosmic evolution.\\
\noindent \textbf{Pressure Component with Viscosity} Analyzes how viscosity influences the pressure, potentially altering the universe's expansion behavior. Viscosity can play a role in dissipating energy, influencing the overall expansion rate or cosmic evolution.

In this section, your analysis focuses on the pressure component of the cosmic fluid, which includes the effects of viscosity, a crucial factor influencing the expansion and evolution of the universe.
\begin{figure}[H]
	\centering
	\includegraphics[scale=0.7]{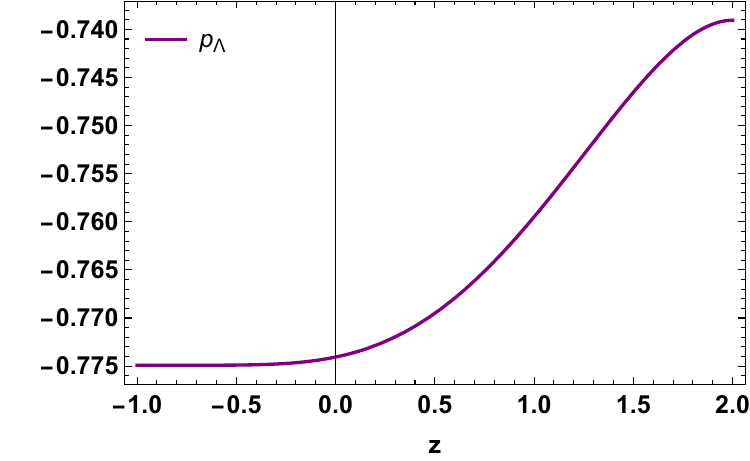}
	\caption{The behavior of isotropic pressure of the fluid with the values of free parameters constraint by the combined data sets of DESI-Y1, SDSS-IV, Pantheon$+$ (without SHOES) and CC.}\label{p}
\end{figure}
Fig. (\ref{p}) highlights that the pressure component remains negative across all redshift values. This negative pressure is significant because it directly contributes to the accelerating expansion of the universe. Negative pressure is a common characteristic of dark energy (DE), which is believed to drive the universe's accelerated expansion. Bulk viscosity arises when a fluid resists compression, and in a cosmological context, it can exert repulsive gravitational effects. This repulsive effect counters the gravitational pull of matter, leading to an accelerated expansion of the universe. The presence of viscosity modifies the cosmic fluid's pressure, enabling it to exhibit properties similar to dark energy. This makes bulk viscosity a potential mechanism to explain cosmic acceleration without requiring an explicit dark energy component. The behavior of the bulk viscous fluid in our viscous $f(Q, C)$ cosmological model mirrors that of a DE component, where negative pressure is essential to explaining the universe's expansion at an accelerating rate. This suggests that bulk viscosity could be a viable alternative or complement to the dark energy hypothesis in explaining the observed dynamics of the universe.\\ 
The negative pressure behavior seen in the bulk viscous cosmic fluid provides strong support for the idea that viscosity can play a crucial role in the universe's expansion. This reinforces the potential of your viscous $f(Q, C)$ cosmological model to offer a theoretical foundation for cosmic acceleration, aligning it with observed phenomena while potentially reducing the reliance on a separate dark energy component. By modeling the universe's cosmic fluid with bulk viscosity, your study contributes to the growing body of work that explores alternative mechanisms for explaining the universe's accelerated expansion, which remains one of the central challenges in modern cosmology.\\

\begin{figure}[H]
		\centering
		\includegraphics[scale=0.7]{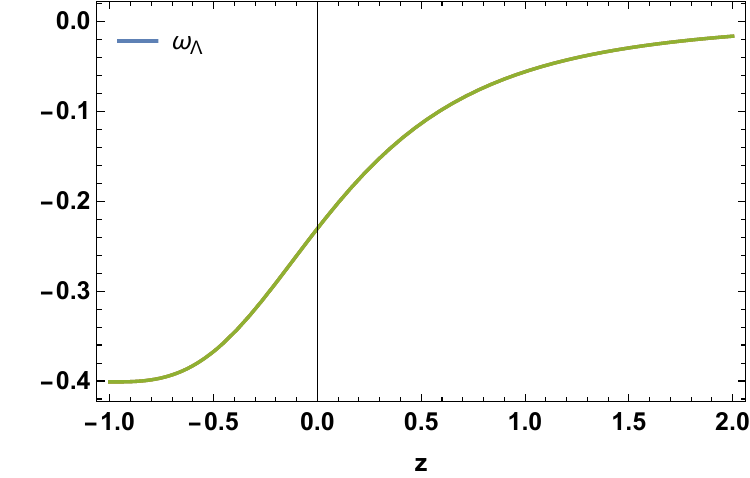}
		\caption{The behavior of equation of state parameter of dark energy fluid with the values of free parameters constraint by the combined data sets of DESI-Y1, SDSS-IV, Pantheon$+$ (without SHOES) and CC.}\label{eosd}
	\end{figure}

\noindent \textbf{Equation of State (EoS) Parameter:} Here we investigated both equation of state parameters namely equation of state parameter for dark energy $\omega_{\Lambda}$ and for viscous fluid $\omega_{\text{eff}}$, which provides a ratio of pressure to energy density. The effective equation of state parameter is critical for understanding how different phases of the universe's expansion (e.g., acceleration or deceleration) unfold over time.\\
The dark energy equation of state parameter:

The plot in the Fig. (\ref{eosd}) of the equation of state (EoS) parameter, provides insights into how dark energy behaves over different epochs in cosmic history. When $z > 0$, the $\omega_\Lambda$ is positive, indicating that dark energy behaves more like conventional matter with minimal repulsive or attractive effect. This suggests that in earlier cosmic times, dark energy may have had a weaker influence on cosmic expansion, allowing matter to dominate the dynamics of the universe. As redshift decreases and approaches $z = 0$ (near the present), it shifts to negative values, around $-0.22$. This transition marks the era where dark energy becomes more dominant, exerting a significant repulsive effect. A negative $\omega_\Lambda$ value implies that dark energy is now pushing the universe to expand at an accelerated rate. At $z < 0$, representing the future state of the universe, the EoS parameter reaches approximately $-0.4$. This moderately negative value suggests that dark energy is contributing to the accelerated expansion but is less extreme than in models where 
$\omega_\Lambda$ would be closer to $-1$, like the cosmological constant ($\Lambda$)  in $\Lambda$CDM. This value hints that dark energy in this model does not behave exactly like a cosmological constant but rather has a slightly evolving or dynamic nature.\\
However, the curve indicates that dark energy transitioned from a negligible, almost matter-like influence in the past to a repulsive, expansion-driving force at present, with the current EoS value around $-0.4$, indicating a dynamic form of dark energy rather than a fixed cosmological constant.\\

\noindent \textbf{The effective equation of state parameter:}

\begin{figure}[H]
		\centering
		\includegraphics[scale=0.7]{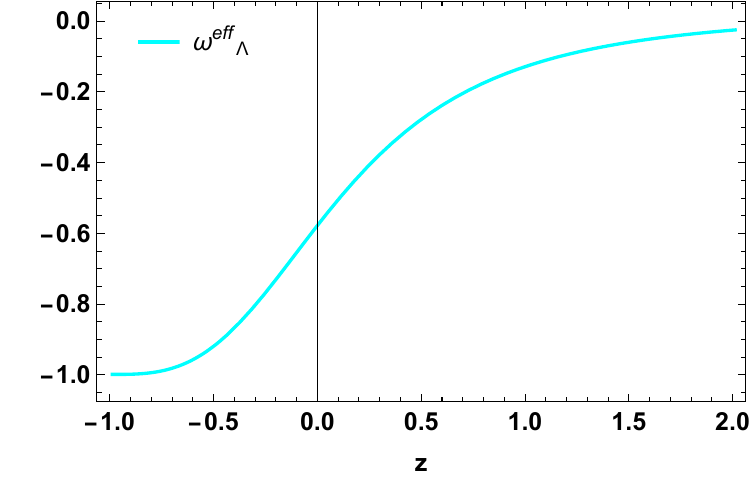}
		\caption{The behavior of equation of state parameter of viscous dark energy fluid with the values of free parameters constraint by the combined data sets of DESI-Y1, SDSS-IV, Pantheon$+$ (without SHOES) and CC.}\label{eosv}
\end{figure}

The analysis of the effective equation of state (EoS) parameter in Fig. \ref{eosv} reveals important insights into the evolution of the universe's cosmic fluid, especially under the influence of viscosity. The EoS parameter, denoted as $\omega_\Lambda^{eff}$, plays a key role in characterizing the different phases of the universe's expansion. The cosmic viscous fluid begins in a matter-dominated phase, where the effective EoS parameter $\omega_\Lambda^{eff}=0$. This reflects a period in cosmic history where the pressure was negligible compared to the energy density, similar to non-relativistic matter (dust), which dominates the universe's expansion at early times. As the universe evolves, $\omega_\Lambda^{eff}$ crosses into the quintessence region, characterized by $-1< \omega_\Lambda^{eff}<0$. Quintessence refers to a dynamic form of dark energy where the EoS is less than zero but greater than \( -1 \), indicating a universe driven by a repulsive force that is still weaker than a cosmological constant. This transition highlights the gradual shift in dominance from matter to a more "fluid-like" dark energy component. \\


The present value of the effective EoS parameter, derived from combining datasets like $H(z)$ and $Pantheon$, is around $\omega_0 \approx -0.60$. This indicates that the universe is currently in an accelerating expansion phase but still in the quintessence-like regime, where $\omega_\Lambda^{eff}$ has not yet reached the value \( -1 \) associated with the cosmological constant in the \( \Lambda\text{CDM} \) model. This intermediate value of \( \omega_0 \) suggests that the universe is transitioning toward a dark energy-dominated future but has not yet reached the full cosmological constant-driven expansion. The current expansion is faster than it was in the matter-dominated era but still distinct from the late-time exponential expansion predicted by \( \Lambda\text{CDM} \). In the far future, the effective EoS parameter approaches $\omega_\Lambda^{eff}=-1$, which corresponds to the $ \Lambda$CDM model. This model describes a universe dominated by a cosmological constant \( \Lambda \), which is responsible for the current phase of accelerated expansion. In this phase, dark energy (modeled as a cosmological constant) drives a continuous, exponential expansion of space.\\
The evolution of the effective EoS parameter described in Fig. (\ref{eosv}) provides a comprehensive picture of the universe's dynamic expansion. It captures the transition from a matter-dominated era to a quintessence-driven phase and finally to the cosmological constant-dominated future. The current value of $\omega_0 \approx -0.60$ highlights that the universe is undergoing accelerated expansion but with characteristics akin to quintessence, rather than a pure cosmological constant. This behavior, consistent with observational data, reinforces the potential of the viscous $f(Q, C)$ cosmological model to explain the universe's expansion history and its current state, offering a plausible framework that aligns with both the standard $\Lambda\text{CDM}$ model and the notion of evolving dark energy.\\

\noindent \textbf{Total density parameter} The total density parameter \( \Omega \) is a key quantity in cosmology, representing the ratio of the total energy density of the universe to the critical energy density, \( \rho_c \). It is typically expressed as the sum of the matter density parameter \( \Omega_m \) and the dark energy density parameter \( \Omega_\Lambda \):
\[
\Omega = \Omega_m + \Omega_\Lambda
\]
The matter density parameter (\( \Omega_m \)), which includes both baryonic matter (ordinary matter such as atoms and stars) and dark matter. It reflects the amount of matter in the universe that exerts gravitational attraction, slowing down the expansion whereas The dark energy density parameter (\( \Omega_\Lambda \)), associated with the energy responsible for the accelerated expansion of the universe. In the standard $\Lambda$CDM model, this is attributed to the cosmological constant $\Lambda$, which represents a uniform energy density permeating space.\\

\begin{figure}[H]
	\centering
	\includegraphics[scale=0.7]{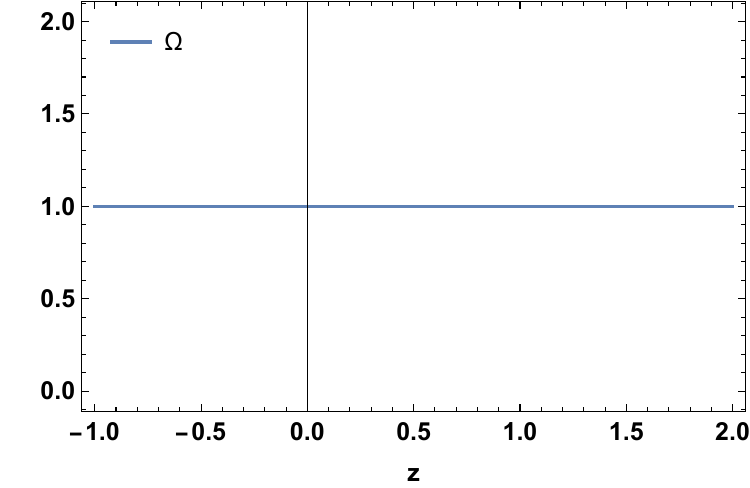}
	\caption{The behavior of total density parameter the fluid with the values of free parameters constraint by the combined data sets of DESI-Y1, SDSS-IV, Pantheon$+$ (without SHOES) and CC.}\label{omega}
\end{figure}

When the total density parameter $\Omega = 1 $, the universe is considered spatially flat. This means that the total energy density of the universe equals the critical density, and the geometry of the universe is flat, as predicted by general relativity. Observations, such as those from the cosmic microwave background (CMB) and galaxy surveys, strongly suggest that the total density parameter is very close to 1. In a flat universe, the sum of the matter and dark energy components satisfies: $\Omega_m + \Omega_\Lambda = 1$. This implies that the universe's total energy density is precisely balanced between matter and dark energy, maintaining a flat geometry. Observational evidence, such as from the Planck satellite, has shown that the total density parameter is indeed very close to 1, with matter (\( \Omega_m \)) contributing roughly 30\%, and dark energy (\( \Omega_\Lambda \)) contributing approximately 70\%.
The value of $\Omega_\Lambda \approx 0.7 $ suggests that dark energy dominates the current universe's energy content, driving its accelerated expansion.
In our derived model the expressions of $\Omega_m$, $\Omega_\Lambda$ and $\Omega$ is obtained as:\\
The total density parameter
\begin{equation}
	\Omega=6^{\alpha -1} (2 \alpha -1) a_1 \left(-H^2\right)^{\alpha -1}
\end{equation}
where $\Omega_m=\frac{\rho_0 (z+1)^{-3 \sigma +3 \omega_m+3}}{3 H^2}$ and $\Omega_\Lambda=\frac{6^{\alpha } (1-2 \alpha ) a_1 \left(-H^2\right)^{\alpha }+2 \rho_0 (z+1)^{-3 \sigma +3 \omega_m+3}}{6 H^2}$. \\Here in the fig. (\ref{omega}), the total density parameter $\Omega = 1$ reflects a universe that is spatially flat and dominated by dark energy, with a significant contribution from matter. This balance between matter and dark energy has profound implications for the universe's geometry, expansion history, and future evolution.

\section{Cosmic kinematical parameters}
\subsubsection{The deceleration parameter}
The deceleration parameter \( q \) is a crucial quantity in cosmology that characterizes the rate of acceleration or deceleration of the universe's expansion. It is defined as:

\[
q = -\frac{\ddot{a}}{a H^2}
\]

where \( \ddot{a} \) is the second derivative of the scale factor \( a(t) \) with respect to time, representing the acceleration of the universe. \( H \) is the Hubble parameter, which measures the rate of expansion of the universe at a given time. \( a(t) \) is the scale factor that describes how distances in the universe change with time. In this analysis it is already defined in equation (\ref{8}). As, we know, if \( q > 0 \), the universe's expansion is slowing down. This would have been the case in earlier epochs dominated by matter or radiation, where gravitational attraction was strong enough to decelerate the universe's expansion. If \( q = 0 \), the universe is expanding at a constant rate, meaning there is no acceleration or deceleration. This is a theoretical condition not commonly observed in standard cosmology and if \( -1 < q < 0 \), the universe is undergoing accelerated expansion. This corresponds to the current epoch, where the expansion of the universe is accelerating due to the influence of dark energy or a similar component. This accelerated expansion is confirmed by observations, such as those of distant supernovae. If \( q = -1 \), the universe undergoes exponential expansion, also known as de Sitter expansion. This is characteristic of a universe dominated by a cosmological constant \( \Lambda \) or dark energy with a constant equation of state \( \omega = -1 \), where the expansion is driven by a uniform repulsive force. The expansion rate does not slow down over time in this scenario whereas if \( q < -1 \), the expansion is super-exponential, meaning that the universe is expanding faster than exponential growth. This occurs in some theoretical models, such as those involving "phantom energy," where the energy density increases over time, leading to a rapid acceleration.
\begin{figure}[H]
\centering
\includegraphics[scale=0.7]{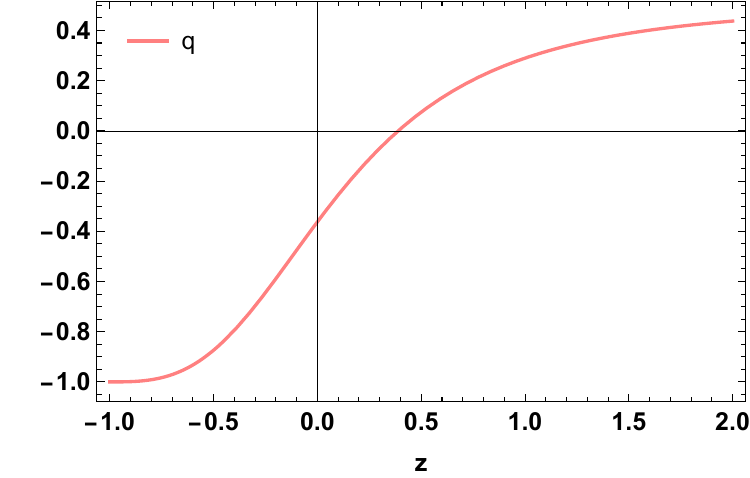}
\caption{The behavior of deceleration parameter of the fluid with the values of free parameters constraint by the combined data sets of DESI-Y1, SDSS-IV, Pantheon$+$ (without SHOES) and CC.}\label{q}
\end{figure}
In this model, the deceleration parameter might evolve differently over time. In earlier epochs, the model could predict a decelerating phase $( q > 0 )$, consistent with a matter-dominated era. As the universe transitions into a dark energy-dominated phase, $q$ could decrease, eventually becoming negative $(q < 0)$, signaling the onset of accelerated expansion. By investigating the behavior of $q$ in this model, the study aims to explore how modifications to gravity and the inclusion of viscosity can influence the expansion rate of the universe, offering potential explanations for the current acceleration and possible deviations from the standard cosmological model.

\subsubsection{The statefinder parameters}

The cosmological constant, while a successful model for explaining the universe's accelerated expansion, faces two major theoretical challenges: 1. \textit{The Cosmological Constant Problem:} This refers to the enormous discrepancy between the observed value of the cosmological constant and the value predicted by quantum field theory. Observationally, $ \Lambda $ is extremely small, while quantum calculations predict a much larger value, leading to a mismatch of around 120 orders of magnitude. 2. \textit{The Cosmic Coincidence Problem:} This problem arises from the fact that the densities of dark energy (represented by $\Lambda$) and matter are currently of the same order of magnitude, even though their evolution over time is very different. This coincidence appears fine-tuned, raising the question of why the universe is in such a balanced state at the present time. To address these challenges, various dynamic models of dark energy have been proposed. In 2003, Sahni et al. introduced a new pair of geometrical parameters, known as statefinder parameters, to distinguish between different dark energy models and provide a tool for comparing them with the standard $\Lambda\text{CDM}$ model. These parameters are designed to go beyond the deceleration parameter $q$ and the Hubble parameter $H$, capturing higher-order characteristics of the universe's expansion. This parameters are defined as follows: 

1. The Statefinder \( r \):
\[
r = \frac{\dddot{a}}{a H^3}
\]
Here, \( \dddot{a} \) represents the third derivative of the scale factor \( a(t) \) with respect to time. This parameter characterizes the rate of change of acceleration, providing information about the "jerk" of the universe's expansion.

2. The Statefinder \( s \):
\[
s = \frac{r - 1}{3(q - 1/2)}
\]
The statefinder parameters $r$ and $s$ are useful tools in cosmology for distinguishing between different models of dark energy. In the standard $\Lambda$-CDM (Lambda Cold Dark Matter) model, these parameters take fixed values: $r=1$ and $s=0$. This makes the $\Lambda$-CDM model a reference point for evaluating other dark energy models, especially those with dynamic components that might deviate from a simple cosmological constant.

\begin{figure}[H]
	\centering
	\includegraphics[scale=0.7]{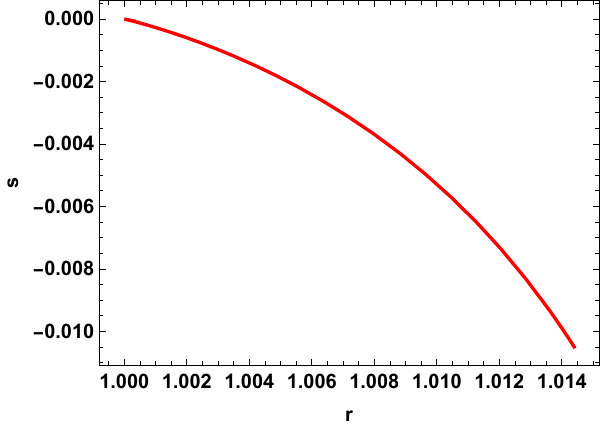}
	\caption{The behavior of statefinder parameters of the fluid with the values of free parameters constraint by the combined data sets of DESI-Y1, SDSS-IV, Pantheon$+$ (without SHOES) and CC.}\label{rs}
\end{figure}

 For example:
\begin{itemize}
	\item Typically, Quintessence models have \( r < 1 \) and \( s > 0 \). Variations in rr and ss in quintessence models help distinguish them from a pure cosmological constant by reflecting the temporal evolution of dark energy (Tsujikawa (2013)).
	\item Models of Phantom energy may show \( r > 1 \) and \( s < 0 \). Phantom dark energy, with an equation of state $\omega < -1$, leads to a more accelerated expansion than $\Lambda$-CDM. This model yields different trajectories for $r$ and $s$ than those of $\Lambda$-CDM, often reflecting divergent behavior over time. Phantom models with dynamical $\omega$-parameters show $r$-values that typically decrease as the universe evolves, and $s$-values may diverge from zero, which helps in distinguishing these models from other dark energy formulations  \textcolor{blue}{(Caldwell (2003))}.
	\item Chaplygin Gas Models: The Chaplygin gas model unifies dark matter and dark energy and is described by an exotic equation of state that transitions from matter-like to dark-energy-like behavior. In this model, the statefinder parameters rr and ss change significantly over time, as the gas undergoes a transition from a matter-dominated era to a dark-energy-dominated era. This dynamic change allows cosmologists to observe how the model deviates from $\Lambda$-CDM across different cosmic epochs  \textcolor{blue}{(Kamenshchik (2001))}.
	\item K-essence Models: K-essence models introduce a kinetic term that drives the evolution of dark energy, leading to distinct predictions for the statefinder parameters. These models can display a range of rr and ss values that shift over time, diverging from  $\Lambda$-CDM by producing different values of the equation of state $\omega$. The evolution of $r$ and $s$ in k-essence models offers a unique approach to modeling dark energy as an evolving field  \textcolor{blue}{(Armendariz-Picon (2001))}.
	
	\end{itemize}
From the evaluated expressions of the statefinder pair $\{r,s\}$, we plot their evolution in Fig.~\ref{rs}. It is observed that the trajectory lies in the vicinity of the $\Lambda$CDM fixed point $\{r,s\}=\{1,0\}$. Specifically, $r$ remains slightly above unity in the range $1.000 \lesssim r \lesssim 1.014$, while $s$ takes small negative values in the interval $-0.01 \lesssim s \lesssim 0$. This behavior indicates that the present viscous $f(Q,C)$ cosmology closely mimics the $\Lambda$CDM model but introduces subtle deviations that can be attributed to the dynamical nature of dark energy in this framework. The slightly positive shift in $r$ together with the negative $s$ suggests a quintessence-like scenario, where the dark energy equation of state evolves dynamically rather than remaining constant as in the case of a cosmological constant. 
\paragraph*{} Therefore, the $\{r,s\}$ diagnostic confirms that the model is consistent with observational expectations while also providing a distinct signature that differentiates viscous $f(Q,C)$ cosmology from the standard $\Lambda$CDM paradigm.

\subsection{The age of the universe}
The age of the universe is obtained as
\begin{align}
\label{age-1}
& dt = -\frac{dz}{(1+z)H(z)}\Rightarrow 
\int_{t}^{t_{0}} dt  =  \int_{0}^{z}\frac{1}{(1+z`)H(z`)}dz`
\end{align}
where $t_{0}$ denotes the age of the universe at present epoch.\\
Eqs. (\ref{9}) and (\ref{age-1}) leads
\begin{equation}
\label{age-2}
H_0 (t_0-t) = \int_{0}^{z}\frac{dz`}{(1+z`) h(z`)}
\end{equation}
where $h(z`) = \frac{H(z`)}{H_{0}}$.\\
\begin{figure}[H]
\centering
\includegraphics[scale=0.70]{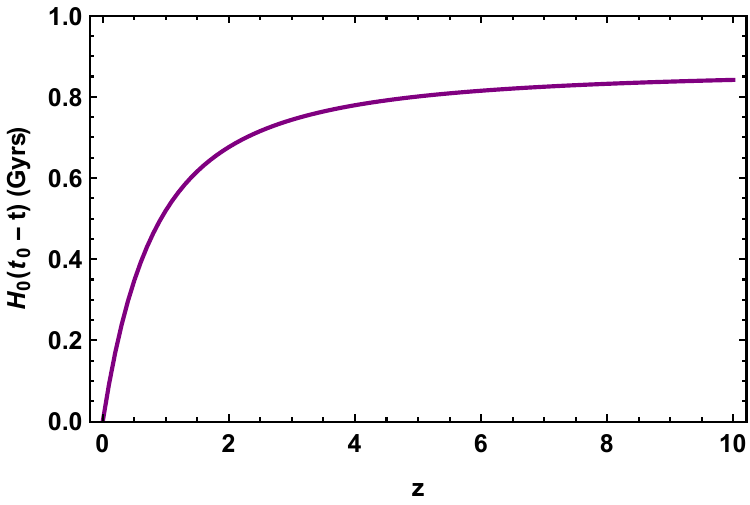}  
\caption{Plot of $H_{0}(t_{0} - t)$ versus redshift $z$.}\label{F10}
\end{figure}
Fig. \ref{F10} shows the graphical representation of $H_{0}(t_{0} - t)$ as a function of redshift $z$. Further, for infinitely high value of $z$, we compute $H_{0}t_{0} \sim 0.87$ which in turn implies that $t_{0} \sim 0.87 H_{0}^{-1}$. Thus, we estimate here the present age of the universe $t_{0} = 13.225 \pm 0.308$ Gyrs. Moreover, the latest Planck collaboration results VI  \textcolor{blue}{(Aghanim (2020))} estimates the present age of universe as $t_{0} = 13.786 \pm 0.020$ Gyrs which confirms that this model estimates age of the universe comparable to the Planck results  \textcolor{blue}{(Aghanim (2020))}.  
\section {Discussion and Conclusion}
In this work, we have explored the cosmological implications of viscous dark energy in the context of $f(Q, C)$ gravity, using a two-fluid approach. The model is designed to go beyond the limitations of $\Lambda$CDM by incorporating non-metricity and boundary terms, which allow for a richer geometrical description of the universe's expansion. By introducing a bulk viscosity term into the dark energy fluid, we have successfully modeled the universe's late-time acceleration without relying on a cosmological constant. 
	
The key finding of this study is that the inclusion of viscosity alters the pressure of the cosmic fluid, leading to negative pressure, which drives the accelerated expansion. This effect is reflected in the behavior of the effective EoS parameter $\omega_{\text{eff}}$, which starts from a value close to zero during the matter-dominated era and decreases as the universe transitions into a phase of accelerated expansion. This transition is consistent with the behavior predicted by dark energy models, such as quintessence or phantom energy, where the EoS parameter varies dynamically over time. We constrained the model parameters using several observational datasets, including DESI-Y1, SDSS-IV, Pantheon+ (without SHOES calibration), and Cosmic Chronometer data. The combined $\chi^2$ minimization technique provided strong constraints on the Hubble parameter $H(z)$, showing excellent agreement with the observational data. The results indicate that the model can accurately reproduce the observed behavior of the Hubble parameter and energy density at different redshifts. Moreover, the model satisfies the physical requirements of energy conditions, such as the null, dominant, and weak energy conditions. These results support the idea that viscous dark energy in the context of $f(Q, C)$ gravity can provide a viable alternative to $\Lambda$CDM while offering additional insights into the universe's dynamics. The analysis of the deceleration parameter and statefinder parameters further emphasizes the model's ability to describe the transition from deceleration to acceleration and distinguishes it from other dark energy models.\\
In summary, the novelty of this work lies in unifying three distinct ingredients-viscous dark energy, the geometric extensions of $f(Q,C)$ gravity, and a two-fluid interaction scenario-into a single framework that can replicate observational features while deviating in testable ways from $\Lambda$CDM. Unlike the standard cosmological constant, the present model predicts a dynamical effective equation of state ($\omega_{\text{eff}} \approx -0.60$ at $z=0$), a statefinder trajectory with $(r,s)$ slightly displaced from $(1,0)$, and a cosmic expansion history driven by both viscosity and non-metricity effects. These distinctive predictions highlight the capacity of viscous $f(Q,C)$ cosmology to provide an alternative, geometrically motivated explanation for late-time acceleration, setting it apart from conventional $\Lambda$CDM.
	
Hence, the viscous $f(Q, C)$ gravity model offers a promising framework for explaining the universe's accelerating expansion. The combination of bulk viscosity and non-metricity provides a flexible and testable approach to cosmological evolution. Future research could further refine the model by incorporating more precise observational data and exploring additional interactions between dark matter and dark energy. It will be part of our next study strategy.

\section*{Acknowledgments}
We appreciate the editor and the anonymous reviewers for their insightful suggestions which greatly improved the manuscript. The IUCAA, Pune, India, is acknowledged by the authors (A. Pradhan and S. H. Shekh) for providing facilities under the Visiting Associateship program. Additionally, the Science Committee of the Republic of Kazakhstan's Ministry of Science and Higher Education provided funding for the research (Grant No. AP23483654).

\section*{Author contributions}
\noindent
\textbf{N. Myrzakulov} --	Conceptualization, Methodology, Supervision, Writing – Review \& Editing.\\
\textbf{Anirudh Pradhan}	--	Investigation, Formal Analysis, Data Curation, Writing – Original Draft.\\
\textbf{S. H. Shekh} --	Methodology, Validation, Formal Analysis, Visualization, Supervision, Writing – Review \& Editing.\\
\textbf{Anil Kumar Yadav} --	Funding Acquisition, Resources, Writing – Review \& Editing, Project Administration

\section*{Funding}
The authors received no specific funding for this work.

\section*{Data availability}
No data was used for the research described in the article.

\section*{Declaration of competing interest}
The authors declare no competing interests.

\section*{Ethical declaration}
The authors declare no ethical declaration.

.

\section{References}

Ade, P.A.R., Aghanim, N., Arnaud, M., Ashdown, M., Aumont, J., et al., Planck 2015 results. XIII. Cosmological parameters, Astron. Astrophys. \textbf{594} (2015) A13.\\

Aghanim, N., Akrami, Y., Ashdown, M., Aumont, J., Baccigalupi, C., et al., Planck 2018 results-vi. Cosmological parameters, Astron. Astrophys. \textbf{641} (2020) A6.\\

Anagnostopoulos, F.K., Basilakos, S., Saridakis, E.N., Observational Constraints on f(Q,T) gravity, Eur. Phys. J. C \textbf{80} (2020) 826.\\

Armendariz-Picon, C., Mukhanov, V., Steinhardt, P.J., A dynamical solution to the problem of a small cosmological constant and late-time cosmic acceleration, Phys. Rev. Lett. \textbf{85} (2000) 4438.\\

Armendariz-Picon, C., Mukhanov, V., Steinhardt, P.J., Essentials of k-Essence, Phys. Revi. D \textbf{63} (2001) 103510.\\

Bamba, K., Capozziello, S., Nojiri, S., Odintsov, S.D., Dark energy cosmology: the equivalent description via different theoretical models and cosmography tests, Astrophys. Space Sci. \textbf{342} (2012) 155.\\

Bengaly, C., Bohringer, H., de Mello, D., Giani, L., The eROSITA Final Equatorial-Survey Galaxy Cluster Catalogue, Eur. Phys. J. C \textbf{83} (2023) 548.\\

Bennett, C.L., Halpern, M., Hinshaw, G., Jarosik, N., Kogut, A., et al., First-Year Wilkinson Microwave Anisotropy Probe (WMAP) observations: Preliminary maps and basic results, The Astrophy. Jour. Suppl. Ser. \textbf{148} (2003) 1.\\

Bento, M.C., Bertolami, O., Sen, A.A., Generalized Chaplygin gas, accelerated expansion, and dark-energy-matter unification, Phys. Rev. D \textbf{66} (2002) 043507.\\

Blanton, R.M., Burles, S.M., Eisenstein, D.J., Gunn, J.E., Knapp, G.R., et al., The Sloan Digital Sky Survey IV, Astron. J. \textbf{154} (2017) 35.\\

Bolotin, Y.L., Kostenko, A., Lemets, O.A., Yerokhin, D.A., Cosmological evolution with interaction between dark energy and dark matter, Int. J. Mod. Phys. D \textbf{24} (2015) 1530007.\\

Brevik, I., Gr\o{}n, \O{}. , de Haro, J., Odintsov, S.D., Saridakis, E.N., Viscous Cosmology for Early- and Late-Time Universe, Int. J. Mod. Phys. D \textbf{26} (2017) 1730024.\\

Caldwell, R.R., Doran, M., Dark-energy evolution across the cosmological-constant boundary, Phys. Rev. D \textbf{69} (2004) 103517.\\

Caldwell, R.R., Kamionkowski, M., Weinberg, N.N., Phantom energy and cosmic doomsday, Phys. Rev. Lett. \textbf{91} (2003) 071301.\\

Capozziello, S., De Falco, V., Ferrara, C., The role of the boundary term in f(Q, B) symmetric teleparallel gravity, Eur. Phys. J. C \textbf{83} (2023) 915.\\

Capozziello, S., Faraoni, V., Beyond Einstein Gravity: A Survey of Gravitational Theories for Cosmology and Astrophysics, Springer (2011).\\

Caprini, C., Tamanini, N., Constraining early and interacting dark energy with gravitational wave standard sirens: the potential of the eLISA mission, JCAP \textbf{10} (2016) 006.\\

Daniel, S.F., Constraining scalar-tensor theories of gravity through cosmology, Phys. Rev. D \textbf{77} (2008) 103513.\\

De, A., Loo, T.-H., Saridakis, E.N., Non-metricity with bounday terms: f(Q,C) gravity and cosmology, JCAP \textbf{2024} (2024) 050.\\

D'Ambrosio, F., Garg, M., Heisenberg, L., Nonlinear extension of Non-Metricity scalar for MOND, Phys. Lett. B \textbf{797} (2019) 134848.\\

Dialektopoulos, K.F., Dunsby, P.K.S., Variational principle for the f(Q) gravity theory, Phys. Rev. D \textbf{103} (2021) 043509.\\

Eisenstein, D.J., Blanton, M., Burles, S., Eisenstein, D.J., Gunn, J.E., et al., Detection of the baryon acoustic peak in the large-scale correlation function of SDSS luminous red galaxies, The Astrophys. Jour. \textbf{633} (2006) 560.\\

Elizalde, E., Nojiri, S., Odintsov, S.D., Late-time cosmology in (phantom) scalar-tensor theory: Dark energy and the cosmic speed-up, Phys. Rev. D \textbf{70} (2004) 043539.\\

Ferraro, R., Fiorini, F., Modified teleparallel gravity: Inflation without an inflaton, Phys. Rev. D \textbf{75} (2007) 084031.\\

Frusciante, N., Signatures of f(Q,C) gravity in cosmology, Phys. Dark Univ. \textbf{33} (2021) 100847.\\

Harko, T., Lobo, F.S.N., et al., \emph{f(Q,T) gravity}, Phys. Rev. D \textbf{98} (2018) 084043.\\

Heisenberg, L., Review on f(Q) gravity, arXiv:2309.15958[gr-qc] (2023).\\

Hinshaw, G., Larson, D., Komatsu, E., Spergel, D.N., et al., Nine-year Wilkinson Microwave Anisotropy Probe (WMAP) observations: Cosmological parameter results, The Astrophy. Jour. Suppl. Ser. \textbf{208} (2013) 19.\\

Huang, Z., Lu, H.Q., Wang, W., Wang, B., Supernova constraints on models of dark energy revisited, Mon. Not. Roy. Astron. Soc. \textbf{368} (2006) 1252.\\

Ishak, M., Collaboration, D., DESI Year-1 Cosmology Results (Dark Energy, Modified Gravity, and Neutrinos), Bulletin of the AAS \textbf{57} (2025) 2.\\

Jimenez, J.B., Heisenberg, L., Koivisto, T., Coincident General Relativity, Phys. Rev. D \textbf{98} (2018) 044048.\\

Jimenez, J.B., Heisenberg, L., Koivisto, T.S., The Geometrical trinity of gravity, Universe \textbf{5} (2019) 173.\\

Kamenshchik, A., Moschella, U., Pasquier, V., An Alternative to Quintessence, Phys. Lett. B \textbf{511} (2001) 265.\\

Khoury, J., Weltman, A., Chameleon cosmology, Phys. Rev. D \textbf{69} (2004) 044026.\\

Koivisto, T., Mota, D.F., Cosmology and astrophysical constraints of Gauss-Bonnet dark energy, Phys. Rev. D \textbf{73} (2006) 083502.\\

Krssak, M., van den Hoogen, R.J., Pereira, J.G., Bohmer, C.G., Coley, A.A., Teleparallel theories of gravity: illuminating a fully invariant approach, Class. Quantum Grav. \textbf{36} (2019) 183001.\\

Lazkoz, R., Salzano, V., Alcaniz, J.S., Observational constraints of teleparallel dark energy, Phys. Rev. D \textbf{100} (2019) 104027.\\

Maurya, D.C., Modified f(Q,C) gravity dark energy models with observational constraints, Mod. Phys. Lett. A \textbf{39} (2024) 2450034.\\

Maurya, D.C., Quintessence behaviour dark energy models in f(Q,B)-gravity theory with observational constraints, Astronomy and Computing \textbf{46} (2024) 100798.\\

Maurya, D.C., Transit cosmological models in non-coincident gauge formulation of f(Q,C) gravity theory with observational constraints, Gravit. Cosmol. \textbf{30} (2024) 330.\\

Maurya, D.C., Cosmology in non-coincident gauge formulation of f(Q,C) gravity theory, Int. J. Geom. Methods Mod. Phys. \textbf{21} (2024) 2450210.\\

Nojiri, S., Odintsov, S.D., Quantum de Sitter cosmology and phantom matter, Phys. Lett. B \textbf{562} (2003) 147.\\

Nojiri, S., Odintsov, S.D., Modified Gauss-Bonnet Theory as an alternative for dark energy, Phys. Lett. B \textbf{631} (2005) 1.\\

Nojiri, S., Odintsov, S.D., Unified cosmic history in modified gravity: from F(R) theory to Lorentz non-invariant models, Phys. Rept. \textbf{505} (2011) 59.\\

Nojiri, S., Odintsov, S.D., Oikonomou, V.K., Modified gravity theories on a nutshell: Inflation, Bounce and Late-time evolution, Phys. Rept. \textbf{692} (2017) 1.\\

Nunes, R.C., Pan, S., Saridakis, E.N., New constraints on interacting dark energy from cosmic chronometers, Phys. Rev. D \textbf{94} (2016) 023508.\\

Padmanabhan, T., Accelerated expansion of the universe driven by tachyonic matter, Phys. Rev. D \textbf{66} (2002) 021301 (2002).\\

Percival, W.J., Reid, B.A., Eisenstein, D.J., et al., Baryon acoustic oscillations in the Sloan Digital Sky Survey Data Release 7 galaxy sample, Mon. Not. Roy. Astron. Soc. \textbf{401} (2010) 2148.\\

Perlmutter, S., Aldering, G., Goldhaber, G., Knop, R.A., Nugent, P., et al., Measurements of $\Omega$ and $\Lambda$ from 42 high-redshift supernovae, Astrophys. Jour. \textbf{517} (1999) 565.\\

Pradhan, A., Dixit, A., Zeyauddin, M., Krishnannair, S., A flat FLRW dark energy model in f(Q,C)-gravity theory with observational constraints, Int. J. Geom. Methods Mod. Phys. \textbf{21} (2024) 2450167.\\

Ratra, B., Peebles, P.J.E., Cosmological consequences of a rolling homogeneous scalar field, Phys. Rev. D \textbf{37} (1998) 3406.\\

Riess, A.G., Filippenko, A.V., Challis, P., Clocchiatti, A., Diercks, A., et al., Observational evidence from supernovae for an accelerating universe and a cosmological constant, Astron. Jour. \textbf{116} (1998) 1009.\\

Salvatelli, V., Said, N., Bruni, M., Melchiorri, A., Wands, D., Phys. Rev. Lett. \textbf{113} (2014) 181301.\\

Samaddar, A., Singh, S.S., Muhammad, S., Zotos, E.E., Behaviours of rip cosmological models in f(Q,C) gravity, Nucl. Phys. B \textbf{1006} (2024) 116643.\\

Sami, M., Toporensky, A., Phantom field and the fate of the universe, Mod. Phys. Lett. A \textbf{19} (2004) 1509.\\

Scolnic, D., et al., The Pantheon+ Analysis: Supernova Constraints on the Hubble Constant and Dark Energy, ApJ \textbf{938} (2022) 113.\\

Sotiriou, T.P., Faraoni, V., f(R) theories of gravity, Rev. Mod. Phys. \textbf{82} (2010) 451.\\

Spergel, D.N., Verde, L., Peiris, H.V., Komatsu, E., Nolta, M.R., et al., First-Year Wilkinson Microwave Anisotropy Probe (WMAP) observations: Determination of cosmological parameters, The Astrophy. Jour. Suppl. Ser. \textbf{148} (2003) 175.\\

Tsujikawa, S., Quintessence: A Review, Class. Quantum Grav. \textbf{30} (2013) 214003.\\

Usman, M., Jawad, A., Sultan, A.M., Compatibility of gravitational baryogenesis in f(Q, C) gravity, Europ. Phys. J. C \textbf{84} (2024) 868.\\

Wang, B., Abdalla, E., Barandela, F.A., Pavon, D., Dark Matter and Dark Energy Interactions: Theoretical Challenges, Cosmological Implications and Observational Signatures, Rept. Prog. Phys, \textbf{79} (2016) 096901.\\

Xia, D.-M., Wang, S., Constraining interacting dark energy models with latest cosmological observations, Mon. Not. Roy. Astron. Soc. \textbf{463} (2016) 952.\\

Zarrouki, R., Bennai, M., Dynamical aspects of scalar phantom field in DGP cosmological model, Int. Jour. Mod. Phys. A \textbf{25} (2010) 2507.\\

Zhao, D., Luo, X., Cosmological constraints on f(Q) gravity: Deviations from $\Lambda$CDM, Phys. Dark Univ. \textbf{35} (2022) 100994.\\

Zhao, W., Cai, Y., Cosmological dynamics of f(Q,C) gravity, Jour. Cosmol. Astrop. Phys. \textbf{2021} (2021) 034.\\

Jimenez R. and Loeb, A. (2002). Constraining cosmological parameters based on relative galaxy ages. Astrophysical Journal, 573, 37.

Moresco, M. (2012). Improved constraints on the expansion rate of the Universe up to z= 1.1 from the spectroscopic evolution of cosmic chronometers. Journal of Cosmology and Astroparticle Physics, 2012(08), 006.

Moresco, M., et al. (2016). A 6\% measurement of the Hubble parameter at z ~0.45: direct evidence of the epoch of cosmic re-acceleration. Journal of Cosmology and Astroparticle Physics, 2016(05), 014.

Moresco, M., et al. (2022). Unveiling the Universe with cosmic chronometers. I. New constraints on the Hubble parameter with cosmic chronometers. Astrophysical Journal, 935, 20.

Scolnic, D. M., et al. (2018). The complete light-curve sample of spectroscopically confirmed SNe Ia from Pan-STARRS1 and cosmological constraints from the combined Pantheon sample. Astrophysical Journal, 859, 101.

Brout, D., et al. (2022). The Pantheon+ Analysis: Cosmological Constraints. Astrophysical Journal, 938, 110.

Adame A. G., DESI Collaboration (2024). DESI 2024 VI: Baryon Acoustic Oscillations from the First Year of DESI. arXiv:2404.03000.

Foreman-Mackey, D., Hogg, D. W., Lang, D., \& Goodman, J. (2013). emcee: The MCMC Hammer. Publications of the Astronomical Society of the Pacific, 125, 306.

\end{document}